\documentstyle[pre,aps,multicol,psfig]{revtex}

\begin{document}
\draft
\title{Universality and multifractal behaviour of spin-spin
	correlation functions in disordered Potts models}
\author{Christophe Chatelain and Bertrand Berche}
\address{Laboratoire de Physique des Mat\'eriaux,\footnote{Unit\'e
	Mixte de Recherche CNRS No~7556} 
	Universit\'e Henri Poincar\'e Nancy 1, \\
	B.P.~239, F - 54506 Vand{\oe}uvre les Nancy Cedex, France}

\date{\today}

\maketitle

\begin{abstract}
We report a transfer matrix study of the random bond $q-$state Potts model in the
vicinity of the Ising model $q=2$. We draw attention to 
a precise determination
of magnetic scaling dimensions in order to compare with perturbative results.
Universality is checked by the computation of the spin-spin correlation function 
decay exponent obtained with different types of probability distributions of
the coupling strengths. Our numerical data, compared to 
perturbative results for the second 
moment of the correlation
function, obtained with both replica symmetry
and replica symmetry breaking schemes, are conclusively in favour of the replica
symmetric calculations.
The multifractal behaviour of higher moments as well as that of typical
correlation functions are also investigated and a comparison is made with the
perturbative expansions. Finally, the shape of the correlation function probability
distribution is
analyzed. 
\end{abstract}



\begin{multicols}{2} 
\narrowtext

\section{Introduction}
\label{sec:intro}

Random systems represent the paradigm of spatially inhomogeneous systems 
where scale invariance is only preserved on average, but not for specific
disorder realizations~\onlinecite{dotsenko95}. In such systems, not a single
exponent but instead an infinite hierarchy of independent exponents are
expected to describe the scaling behaviour of local quantities like 
order parameter density 
profiles or correlation functions. This property, linked to the 
non self-averaging behaviour of the corresponding physical 
quantities~\onlinecite{aharonyharris96,derrida84,wisemandomany98a,wisemandomany98} is 
ususally referred to as multifractality. 
The keystone concept which enters the description of multiscaling
properties is that of scaling dimensions associated to the moments of the 
local physical property, or equivalently the universal function $H(\alpha)$ corresponding to the Legendre 
transform of the set of independent scaling indexes (See e.g. 
Refs.~\onlinecite{stanleymeakin88,janssen94}).

Ten years ago already, in a series of illuminating papers, Ludwig~\onlinecite{ludwig87,ludwig90}
and Ludwig and Cardy~\onlinecite{ludwigcardy87} reported an extensive analytic 
study of $2D$ random bond Potts ferromagnets in the regime where bond randomness
is slightly relevant, $q$ close to 2, $q$ being the number of states per spin.
Their studies included perturbative 
calculations of the conformal anomaly, of the thermal scaling dimension and 
of the
multifractal behaviour of spin-spin correlation functions.

The essential
of the numerical studies dealing with scaling dimensions of average quantities 
in random bond Potts models at small
values of $q$ were performed by
Monte Carlo (MC) simulations combined
to standard Finite Size 
Scaling (FSS)~\onlinecite{wisemandomany95,picco96,kim96,picco98}
or transfer matrix (TM) calculations associated to 
conformal methods~\onlinecite{jacobsencardy98,glaus87,picco97,cardyjacobsen97,chatelainberche99}
at the 
random fixed point (FP) of self-dual disordered models.
In particular, an excellent quantitative agreement for the magnetic
scaling dimension in the three-state Potts model was reported in
Refs.~\onlinecite{jacobsencardy98,cardyjacobsen97}.
In which concerns the multiscaling properties of spin-spin correlation
functions, 
although some of Ludwig's predictions have partially been verified both 
in cylinder 
geometry~\onlinecite{jacobsencardy98,palagyichatelainbercheigloi99}
and in square geometry~\onlinecite{olsonyoung99},  
the agreement with analytical expansions was less conclusive and
in particular
the shape of the probability distribution has not been reproduced.

Monte Carlo simulations are not convenient to study numerically the vicinity of 
the Ising model, $q=2$, where perturbation expansions are supposed to apply. 
The number
of states per spin, $q$, is indeed restricted to integer values in such 
simulations.
In this paper, we therefore use a rather different approach already
used by different 
authors~\onlinecite{jacobsencardy98,cardyjacobsen97,chatelainberche99,chatelainberche98b}. 
This technique is based on the Fortuin-Kasteleyn graph 
representation~\onlinecite{fortuinkasteleyn69} which enables
TM calculations~\onlinecite{blotenightingale82} where  $q$ 
enters as a parameter that can take non integer
values. 
We also benefit from previous studies with a bimodal probability distribution 
of spin-spin interactions where the disorder amplitude was found to 
have a deep influence on the measured critical properties in numerical 
studies~\onlinecite{picco98}.
It should thus be chosen carefully in order to avoid crossover perturbations 
due to 
the unstable pure model fixed point, $r=1$, and the percolation fixed point,
$r\to\infty$, $r$ being the ratio between strong and weak couplings.

In this paper, we are mainly interested in the multiscaling behaviour of the
spin-spin correlation functions. The first section reminds the reader of the
essential relevant theoretical results which have been obtained by several groups
using perturbative
techniques around the pure models conformal field theories.
Section~\ref{sec:model} explains the methodology and section~\ref{sec:results}
gives the numerical results: 
\begin{description}
\item[i)] Universality in quenched disordered
ferromagnetic Potts models is
checked using different types of probability distributions of nearest-neighbour 
couplings. 
\item[ii)] The numerical study of the decay exponent of the second moment of the spin-spin 
correlation function
is then compared to perturbative results in order to test Replica Symmetry
and Replica Symmetry Breaking scenarios. 
\item[iii)] Finally, we study other moments
and deduce the shape of the universal functions $H(\alpha)$ for different 
$q$ values. The probability distribution of spin-spin correlation functions
is also analyzed.
\end{description}
The values of $q$ are chosen in the range $2$ to $4$, where the pure
model exhibits a second order phase transition, with a special attention paid
on the neighbourhood of $q=3$.

\section{Summary of the perturbation results}
\label{sec:summary}

\subsection{The $2D$ random Ising model}
\label{subsec:spinspin}

According to the celebrated Harris criterion~\onlinecite{harris74}, quenched randomness
is a marginal perturbation in the $2D$ Ising model. This situation has focused
a considerable interest on the critical properties of the random bond Ising model 
(RBIM), and, after partially conflicting results, disorder was eventually found
to be marginally irrelevant, leading to an unchanged universal behaviour apart
from logarithmic corrections for the ensemble average of some physical 
quantities~\onlinecite{dotsenkodotsenko83,shalaev84,shankar87,shankar88,ludwig88,shalaev94}.
These results were then carefully checked through intensive MC 
simulations~\onlinecite{wangetal90,andreichenkoetal90,talapovshchur94} and
series expansions~\onlinecite{roderadlerjanke98,roderadlerjanke99,mazzeokuhn99}.
We mention that site dilution is still subject to 
controversial interpretations (See e.g. Ref.~\onlinecite{mazzeokuhn99})
in spite of a conclusive recent work leading to the same conclusions
than bond randomness~\onlinecite{plechko98}.

The free energy density in a strip of width $L$ with periodic boundary
conditions was obtained by Ludwig and Cardy~\onlinecite{ludwigcardy87}:
\begin{equation}
	\overline{f(L)}\simeq f_0-\frac{\pi}{6L^2}\left(\frac{1}{2}-128\pi^3\Delta^3
	(1+8\pi\Delta\ln L)^{-3}\right),
\label{f-q=2}
\end{equation}
where $\Delta$, the variance of the Gaussian probability distribution of
exchange interactions, is the
strength of disorder related to the ratio $r$. In this
expression, overbar denotes the
disorder average.
The central charge $c$, defined by the leading size dependence of
the free energy density in the cylinder geometry, $\frac{\pi c}{6L^2}$, thus exhibits logarithmic
corrections which make its exact value $\frac{1}{2}$ difficult to extract 
numerically~\onlinecite{dequeiroz95}.

Using a perturbation expansion, Ludwig later obtained the behaviour of the moments of the spin-spin correlation
function~\onlinecite{ludwig90}:
\begin{equation}
	\overline{\langle\sigma (0)\sigma(\rho)\rangle^ p}
	\simeq\rho^{- p/4}(\Delta\ln\rho)^{ p( p-1)/8}
\label{eq:moments-corr}
\end{equation}
when $\Delta$, the strength of disorder, is strong enough.  Brackets
denote the thermal average.
We can also introduce a reduced
correlation function whose leading power-law behaviour is 
\begin{equation}
	\overline{\langle\sigma (0)\sigma(\rho)\rangle^ p}^{1/ p}\sim
	\rho^{-2x_\sigma}.
\label{eq:reduce-corr}
\end{equation}
Ludwig's results imply that logarithmic corrections are absent in the case of 
the average correlation function ($ p=1$), 
$\overline{\langle\sigma (0)\sigma(\rho)\rangle}\sim
\rho^{-1/4}$, while typical correlation functions ($p=0$) exhibit 
such corrections,
$\exp\overline{\ln \langle\sigma (0)\sigma(\rho)\rangle}\sim\rho^{-1/4}
(\Delta\ln\rho)^{-1/8}$, observed numerically~\onlinecite{dequeirozstinchcombe96}. 
Furthermore, a 
unique scaling dimension $x_\sigma={1\over 8}$ describes the
leading  power-law decay 
of all the moments of the spin-spin correlation function,
$\overline{\langle\sigma (0)\sigma(\rho)\rangle^ p}\sim\rho^{-2 p x_\sigma}$,
and no multiscaling
behaviour is expected apart from the logarithmic correction term.

The surface correlation function has also been studied recently and was found
to be self-averaging~\cite{lajkoigloi99}.

\subsection{The random bond Potts model}
\label{subsec:RBPM}

In the case of the Potts model with $q> 2$, disorder is a relevant perturbation
which modifies the universal critical behaviour and leads to new fixed point 
critical properties.
Studying the effect of a slightly relevant perturbation on the finite-size 
scaling behaviour of the free energy density
in a strip geometry of width $L$ at the new fixed point,
Ludwig and Cardy~\onlinecite{ludwigcardy87}, obtained perturbatively the central 
charge $c'(q)$ of the $2D$ $q-$state Potts model with weak quenched bond 
randomness (here and in the following, primes denote the central charge and the
critical exponents at the disordered fixed point, while unprimed symbols refer
to the pure fixed point quantities).
Using an expansion in $q-2$ around the Ising model, the random anomaly was 
deduced from the random free energy of the strip, 
\begin{equation}
f(L,g^*)=A-\frac{\pi c'(q)}{6L^2}+O(L^{-3}),
\label{eq:E0}
\end{equation}
given in the replica
formalism by the quenched free energy $\partial f(n)/\partial n|_{n\to 0}$:
\begin{equation}
	c'(q)=
	{\frac{1}{2}}\left(1+
	{\frac{7}{4}}y_H-
	{\frac{9}{16}}y_H^2-
	{\frac{5}{64}}y_H^3
	+O(y_H^4)\right),
\label{eq:c_r}
\end{equation}
where $y_H$ is the renormalization group (RG) eigenvalue associated to the bond 
disorder~\onlinecite{harris74}, $y_H=\alpha/\nu=2-2x_\varepsilon(q)$, $x_\varepsilon(q)$ being the
scaling dimension of the energy density in the pure 
model~\footnote{In this paper, we use the notations of 
	Refs.~\onlinecite{ludwig87,ludwig90,ludwigcardy87}: The expansion parameter
	is $y_H$. It is related to the parameter $\epsilon$, linked to the
	deviation from the pure Ising model central charge, used by Dotsenko and 
	co-workers in 
	Refs.~\onlinecite{dotsenkopiccopujol95a,lewis98,dotsenkodotsenkopicco97,dotsenkodotsenkopiccopujol95}
	by $y_H=3\epsilon$.}. 
This latter dimension is obtained for arbitrary $q\le 4$ by the den Nijs 
conjecture~\onlinecite{dennijs79,nienhuis82}, rigourously proved by Dotsenko 
and Fateev~\onlinecite{dotsenkofateev84}:
\begin{equation}
	x_\varepsilon(q)=\frac{1+\mu}{2-\mu},\quad
	0\le
	\mu=\frac{2}{\pi}\cos^{-1}\left({\scriptstyle \frac{1}{2}} \sqrt q\right)
	\le 1,
\label{eq:x_eps}
\end{equation}
and to lowest order, the RG eigenvalue is proportional to $q-2$: 
$y_H=\frac{4}{3\pi}(q-2)+O[(q-2)^2]$. The deviation of the conformal 
anomaly from its pure
fixed point value, 
\begin{equation}
	c(q)=1-\frac{3\mu^2}{2-\mu}, 
\label{eq:cpure}
\end{equation}
is difficult to measure, since it is only of third
order in $y_H$
and it requires
a very good accuracy to distinguish between pure and random 
values.

The thermal exponent was similarly obtained to two-loop order  
by Ludwig~\onlinecite{ludwig87}:
\begin{eqnarray}
	x_\varepsilon'(q)&=&x_\varepsilon(q)+\frac{1}{2}y_H
	+\frac{1}{8}y_H^2+O(y_H^3)\nonumber\\
	&=&1+\frac{1}{8}y_H^2+O(y_H^3),
\label{eq:x_eps-ludwig}
\end{eqnarray}
and the three-loop correction was reported by Jug and 
Shalaev~\onlinecite{jugshalaev}.

The correction to the magnetic exponent requires a three-loop calculation. It
was obtained by Dotsenko et al~\onlinecite{dotsenkopiccopujol95a}
\begin{equation}
	x_\sigma'(q)=x_\sigma(q)+\frac{1}{32}\frac{
	\Gamma^2(-{\scriptstyle\frac{2}{3}})\Gamma^2({\scriptstyle\frac{1}{6}})
	}{
	\Gamma^2(-{\scriptstyle\frac{1}{3}})\Gamma^2(-{\scriptstyle\frac{1}{6}})
	}
	y_H^3+O(y_H^4),
\label{eq:x_s-dotsenko}
\end{equation}
and checked by Picco using MC simulations~\onlinecite{picco96}. In contradistinction 
with the thermal exponent, the deviation from the
pure fixed point value is quite small close to $q=2$.

Searching for multiscaling properties, Ludwig obtained, up to linear order,
the scaling dimension 
of the $p^{th}$-moment~\footnote{In the literature, many different notations have been used for
	the scaling dimensions of the moments of the
	correlation functions. Our notation corresponds to that of 
	Lewis~\onlinecite{lewis99}:
	$x'_{\sigma^p}\leftrightarrow\Delta'_{\sigma^p}$. The 
	correspondence with other works is the following:
	Ludwig~\onlinecite{ludwig90}: $x'_{\sigma^p}\leftrightarrow X_{N}/N$,
	Dotsenko et al~\onlinecite{dotsenkodotsenkopicco97}: 
	$x'_{\sigma^2}\leftrightarrow \Delta'_{\sigma^2}/2$
	and Olson and Young~\onlinecite{olsonyoung99}: 
	$x'_{\sigma^p}\leftrightarrow\eta_n/2$.}\onlinecite{ludwig90} of the reduced spin-spin
correlation function, $\overline{\langle\sigma (0)\sigma(\rho)
\rangle^p}^{1/ p}\sim\rho^{-2x'_{\sigma^p}(q)}$
and Lewis performed recently the computation up to the second 
order~\onlinecite{lewis98,lewis99}
\begin{eqnarray}
	x'_{\sigma^p}(q)&=&x_\sigma(q)-\frac{1}{16}( p-1)y_H\nonumber\\
	&-&\frac{1}{32}( p-1)[A+B( p-2)]y_H^2+O(y_H^3),
\label{eq:x_lewis}
\end{eqnarray}
where $A=\frac{11}{12}-4\ln 2$ and $B=\frac{1}{24}(33-29\sqrt 3\pi /3)$. 
Here, the exponent corresponding to the {\it average} critical correlation
function at the random fixed point is denoted by $x'_{\sigma^1}(q)\equiv
x_\sigma'(q)$, while the {\it typical} behaviour corresponds to $ p=0$.
This
result, obtained in the Replica Symmetry (RS) scenario, contains the special case 
of the second moment performed by 
Dotsenko et al~\onlinecite{dotsenkodotsenkopicco97} in order to compare between
Replica Symmetry 
\begin{eqnarray}
	x'_{\sigma^2}(q)&=&x_\sigma(q)-\frac{1}{16}y_H\nonumber\\
	&+&\frac{1}{32}\left( 	4\ln 2
	-\frac{11}{12}\right)y_H^2+O(y_H^3),
\label{eq:RS}
\end{eqnarray}
and Replica Symmetry Breaking 
(RSB)~\footnote{In Ref.~\onlinecite{dotsenkodotsenkopiccopujol95}, the thermal and magnetic
	exponents have been computed with both RS and RSB scenarios. While 
	Eq.~(\ref{eq:x_s-dotsenko}) for the average behaviour is unchanged 
	up to the third order in the
	RSB scheme, the thermal
	exponent in Eq.~(\ref{eq:x_eps-ludwig}) becomes $x''_\varepsilon(q)=1+
	O(y_H^3)$.}\onlinecite{dotsenkodotsenkopiccopujol95}: 
\begin{eqnarray}
	x''_{\sigma^2}(q)&=&x_\sigma(q)-\frac{1}{16}y_H\nonumber\\
	&+&\frac{1}{32}\left( 	4\ln 2
	-\frac{5}{12}\right)y_H^2+O(y_H^3),
\label{eq:RSB}
\end{eqnarray}

\section{Model and methodolgy}      
\label{sec:model}

In this paper, we consider Potts-spin variables, $\sigma_{j}\in {1,2,\dots,q}$
on the sites of a square lattice
with independent quenched random nearest-neighbour ferromagnetic
interactions $K_{ij}$. These exchange couplings
are taken from a probability distribution ${\cal P}(K_{ij})$, and in most
of our applications, they 
can take two values,  
$K_1=K$, $K_2=rK$, $r>1$ with equal probabilities: 
\begin{equation}
	{\cal P}(K_{ij})=\frac{1}{2}\delta(K_{ij}-K)+
	\frac{1}{2}\delta(K_{ij}-rK),
   	\label{probbin}
\end{equation}
where $r$ measures the strength of disorder. Our methodology will be discussed in this
section with the particular case of probability distribution~(\ref{probbin}) and
the generalization to other distributions will be presented in the next section.   

The Hamiltonian of the model is thus written
\begin{equation}
   	-\beta{\cal H}=\sum_{(i,j)} K_{ij}\delta_{\sigma_{i},\sigma_{j}}.
   	\label{Ham}
\end{equation}

We only consider self-dual models for which the critical point is exactly
known. In the case of the bimodal distribution~(\ref{probbin}), the self-duality point
\begin{equation}
	[\exp(K_c(r))-1][\exp(rK_c(r))-1]=q\;,
\label{duality}
\end{equation}
corresponds to the critical point of the model if only one phase
transition takes place in the system as rigourously shown in
Ref.~\onlinecite{chayesshtengel98}. 

The degree of dilution in the system can be varied by changing the
ratio of the strong and weak couplings, $r$. At $r=1$, one
recovers the perfect $q$-state Potts model, whereas for $r \to \infty$
we are in the percolation limit, where $T_c=0$. The intermediate regime
of dilution $1<r<\infty$ is expected to be controlled by the random
fixed point located at some $r=r^\star(q)$~\onlinecite{chatelainberche99}.
This optimal disorder amplitude, $r^\star(q)$, can be obtained numerically from the maximum
condition of the effective central charge of the disordered system. 
Dotsenko and co-workers for example considered $n$ $q-$state Potts models 
coupled via
energy-energy  interactions and obtained perturbatively the central charge 
deviation from the
decoupling limit (where $c$ is given by the sum of the central charges 
of the decoupled 
models)~\onlinecite{dotsenkojacobsenlewispicco98}: $\Delta c=-\frac{1}{8}\frac{n
(n-1)}{(n-2)^2}y_H^3+O(y_H^4)$. For $n>1$, $\Delta c$ satisfies the 
Zamolodchikov's $c-$theorem according to which there exists a $c-$function
decreasing along RG flows and giving the central charge at the fixed 
point~\onlinecite{zamolo}.
In the case of random systems ($n\to 0$ in the replica approach), the central charge
increases and can be expected to reach a maximum value at an optimal
disorder amplitude where the random FP exponents may be extracted from numerical data.
This
property, linked to non-unitarity in the presence of disorder, is indeed observed in 
simulations~\onlinecite{jacobsencardy98,chatelainberche99,chatelainberche98b}.

In the following we
used a TM technique, based on the Bl\"ote and Nightingale connectivity 
transfer matrix~\onlinecite{blotenightingale82}, which enables to compute the
physical quantities in long cylinders. Since 
transfer operators in the time direction do not commute in disordered systems,
the free energy density is defined by the  leading Lyapunov
exponent. For an infinitely long strip of width $L$
with periodic boundary conditions, the 
leading Lyapunov exponent is given by the Furstenberg
method~\onlinecite{furstenberg63}: 
\begin{equation}
	\Lambda_0(L)=\lim_{m\to\infty}\frac{1}{m}
	\ln\left|\!\left|\left(\prod_{k=1}^m 
	{\bf T}_k\right)
	\mid\! v_0\rangle\right|\!\right|,
\label{eq-Furst}
\end{equation}	
where ${\bf T}_k$ is the transfer matrix and
$\mid\! v_0\rangle$ is a unit initial vector. The quenched  
free energy density
is thus given by 
\begin{equation}
	\overline{f(L)}=-L^{-1}\Lambda_0(L).
\label{eq-F-ave}
\end{equation}	

\begin{figure}[ht]
\centerline{\psfig{file=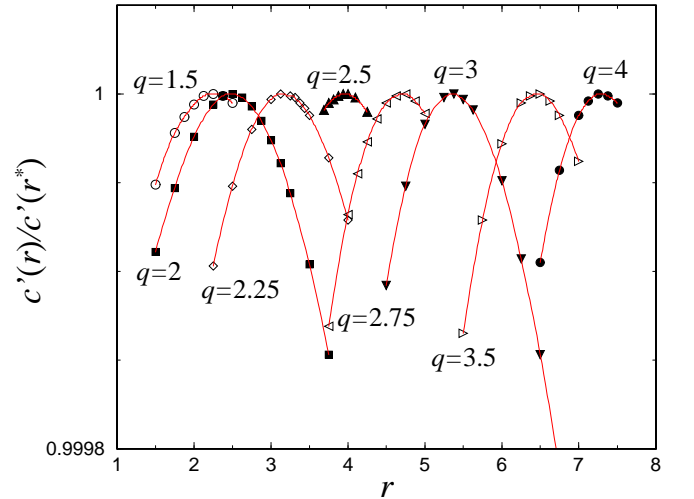,width=8.6cm}}
\vskip 0.2truecm
\caption{Behaviour of the central charge as a function of the disorder 
amplitude for different values of the number of states $q$ (binary distribution
of Eq.~(\protect\ref{probbin})). The solid lines are parabolic fits. The maximum
corresponds to the optimal value of disorder amplitude.} 
\label{fig-max-c} \end{figure}

The shape of the central charge as a function of the disorder 
amplitude is shown in Fig.~\ref{fig-max-c} for several values of $q$. Each realization of disorder is
obtained via $10^6$ iterations of the transfer matrix and the free energy 
density was averaged over 96 such realizations. 

\vbox{
\begin{table}
\caption{Optimal disorder amplitude and corresponding values of the
central charges of the disordered Potts model for different values
of $q$ for the binary probability distribution in 
Eq.~(\protect\ref{probbin}). The last columns give the numerical values of the
pure model central charge $c(q)$ (Eq.~(\protect\ref{eq:cpure})) and of the expansion parameter used in the perturbation results, and shows that the
perturbation expansion obviously breaks down when $q$ increases. \label{table-max-c}}
\begin{tabular}{llllll}
&&\multicolumn{2}{c}{$c'(q)$}\\	
\cline{3-4}
$q$ & $r^*(q)$ & TM result & Eq.~(\ref{eq:c_r}) & $c(q)$ & $y_H$\\
	\hline
1.5	& 2.25 & 0.283(1) &  &0.288& \\
2.	& 2.49 & 0.496(6) & 0.5	&0.500	&   0.\\
2.25	& 3.18 & 0.584(8) & 0.5876&0.588	&   0.1036\\
2.5	& 3.96 & 0.662(8) & 0.6661&0.666	&   0.2036\\	
2.75	& 4.70 & 0.732(8) & 0.7371&0.736	&   0.3017\\	
3.	& 5.36 & 0.797(8) & 0.8020&0.800	&   0.4000\\
3.25	& 5.94 & 0.857(9) & 0.8617&0.858	&   0.5013\\
3.5	& 6.45 & 0.912(11) & 0.9174&0.910	&   0.6101\\
4.	& 7.30	& 1.011(12) & 1.0312&1.000	&   1.\\
\end{tabular}
\end{table}
}

The central charge follows
from a polynomial fit
\begin{equation}
	\overline{f(L)}=f_0-\frac{\pi c}{6L^2}+A_4L^{-4}+A_6L^{-6}+A_8L^{-8},
\label{fit-c}
\end{equation} 
where the remaining coefficients $A_i$ have been included in order to simulate
finite size effects, although the exact dependence of corrections to scaling
is not known. The calculations were performed on strips of
widths $L=2$ to 8.
The central charge is still sensible to the number of disorder
configurations entering the average and to the degree of the polynomial
fit. Nevertheless, it is expected that the position of the maximum
presents small enough deviations for the critical exponents to 
reflect the
disordered fixed point regime in the neighbourhood of this maximum.
We have checked different types of fits, with less parameters and also 
including logarithmic terms
in the vicinity of $q=2$ where equation~(\ref{f-q=2}) is supposed to be valid,
but we were not able to improve the results, since, as it has already been 
observed~\onlinecite{jacobsencardy98}, the values of $c'(q)$ are 
systematically below the perturbative result of
Ludwig and Cardy~\onlinecite{ludwigcardy87}, and even below the pure
model central charge at small values of $q$.  In all the cases, 
the error can
be estimated by the fluctuations of the results with different fitting
procedures and it is of order
$10^{-3}$ to $10^{-2}$.  For example we obtain $c'(2)=0.496$ with Eq.~(\ref{fit-c}),
0.492 with $A_6=A_8=0$,
0.497 with $A_8=0$ but a $(\ln L)^{-3}$ term added or 0.495 with
$A_6=A_8=0$ and the log-term present. The error bars given in 
table~\ref{table-max-c} correspond to the fluctuations of the results obtained
with the different fits.
The maximal values and optimal disorder amplitude $r^*(q)$ are also given in 
table~\ref{table-max-c}.

For a specific disorder realization, the spin-spin correlation function
along the long direction, $u$, of the strip 
\begin{equation}
 	\langle\sigma(j)\sigma(j+u)\rangle=\frac{q\langle\delta_{
	\sigma_j\sigma_{j+u}}\rangle-1}{q-1},
\label{eq-Gu}
\end{equation}	
 where $\langle\dots\rangle$ denotes the 
thermal average, is given by a product of non-commuting transfer matrices.
They were computed on strips of widths $L=2$ to 8 and then averaged over
$80\ 000$ disorder realizations.

We will now assume that
conformal covariance can be applied to the order parameter
correlation function and its moments. 
In the infinite
complex plane $z=x+{\rm i}y$ 
  the correlation function and its moments exhibit the usual
algebraic decay at the critical point
\begin{equation}
	\overline{\langle\sigma(z_1)\sigma(z_2)\rangle^p}^{1/ p}={\rm const}
	\times 
	\rho^{-2 x'_{\sigma^p}(q)}, 
\label{eq:G-plan}
\end{equation}
where
$\rho=\mid\! z_1-z_2\!\mid$. Multiscaling arises when the exponents 
$x'_{\sigma^p}(q)$ are all different, depending upon the value of the 
moment order $ p$, and their $p-$dependence is a convex 
function~\onlinecite{ludwig90}. 
Under the logarithmic transformation
$w=\frac{L}{2\pi}\ln z=u+{\rm i}v$ which maps the infinite plane geometry
inside an infinitely long strip of width $L$, one gets the exponential decay along the 
strip
\begin{equation}
	\overline{\langle\sigma(j)\sigma(j+u)\rangle^p}^{1/ p}={\rm const}\times
	\exp\left[-\frac{2\pi}{L}x'_{\sigma^p}(q)  u\right].
\label{eq:G-ruban}
\end{equation}
The scaling 
dimensions $x'_{\sigma^p}(q)$ at different strip sizes 
can thus be deduced from an exponential fit, and a quadratic extrapolation
at $L\to\infty$ is
performed to get the corresponding value in the thermodynamic limit. The calculation
of errors follows the lines explained in Ref.~\onlinecite{chatelainberche99}.

This method was used in Ref.~\onlinecite{chatelainberche99} for the average correlation
function, i.e. for $ p=1$, and in the large $q$ regime. In this work, we calculate the higher moments, as well as
the typical behaviour, governed by the derivative of the exponent $x'_{\sigma^p}$
with respect to $p$, evaluated in the limit $p\to 0$.
In previous works, it was
shown that the results from exponent extrapolations are sensitive to the value
of the disorder strength chosen for the simulation, since the finite size corrections
are very strong unless the calculations are 
performed close to the random fixed point~\onlinecite{picco98} as  obtained from the maximum 
condition of the central charge of the
model~\onlinecite{chatelainberche99,dotsenkojacobsenlewispicco98}. Our simulations
were performed at these fixed point values of the disorder strength, but for
comparison we have also considered systems with somewhat different values.

\begin{figure}[ht]
\centerline{\psfig{file=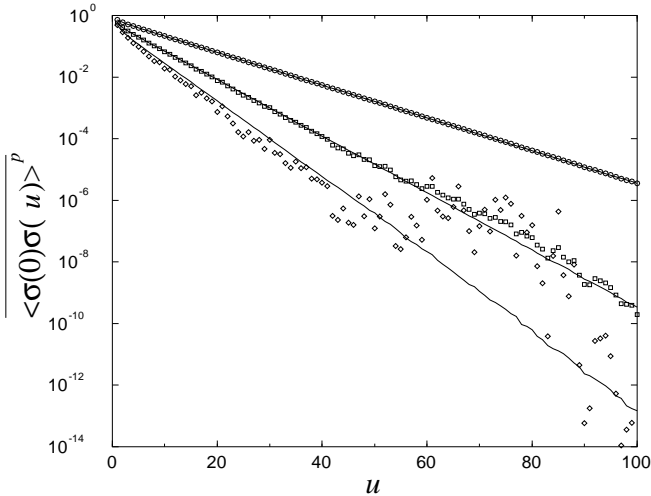,width=8.6cm}}
\vskip 0.2truecm
\caption{Moments of the spin-spin correlation function $p=1$, 2 and 3 from
upper to lower curves ($q=3$, $L=7$, $r=5.36$). The solid lines correpond to the average over 96\ 000 different 
disorder realizations and the symbols are deduced from the cumulant expansion
up to the fifth order. The fluctuations become extremely 
large above the fifth order. Both solid lines and symbols give the same
order for the corresponding scaling dimensions (related to the slopes of these
curves), but the direct average leads to more precise results.} 
\label{fig-cum_Bin} \end{figure}

Since the correlation functions are not self-averaging, 
the disorder average must be performed carefully. 
We follow the same procedure as in 
Ref.~\onlinecite{chatelainberche99} where we compared the
ensemble average $\overline{\langle\sigma(0)\sigma(u)\rangle}$ 
to a cumulant expansion in
terms of the moments of $\ln\langle\sigma(0)\sigma(u)\rangle$, which are
self-averaging~\onlinecite{jacobsencardy98}:
\begin{eqnarray}
	\overline{\langle\sigma(0)\sigma(u)\rangle^p}&=&\exp\left[
	p\ \!\overline{\ln\langle\sigma(0)\sigma(u)\rangle}\right.\nonumber\\
	&+&
	{1\over 2}p^2\left.\left(
	\overline{\ln\langle\sigma(0)\sigma(u)\rangle^2}-
	\overline{\ln\langle\sigma(0)\sigma(u)\rangle}^2
	\right)+\dots\right].
\label{cum-ln}
\end{eqnarray}
Although the average should in principle be done using the cumulant expansion,
we observe that the
direct average, which is compatible with the cumulant expansion, 
is more stable than this latter expansion. This
is particularly true at high moment orders and is
probably due to the large number of disorder 
realizations used in the calculations.
This
is illustrated for several moments at $q=3$ in 
Fig.~\ref{fig-cum_Bin} where the solid lines
represent the direct average over 96\ 000 different samples, while the 
open symbols
correspond to the cumulant expansion up the the fifth order.

\begin{figure}[ht]
\centerline{\psfig{file=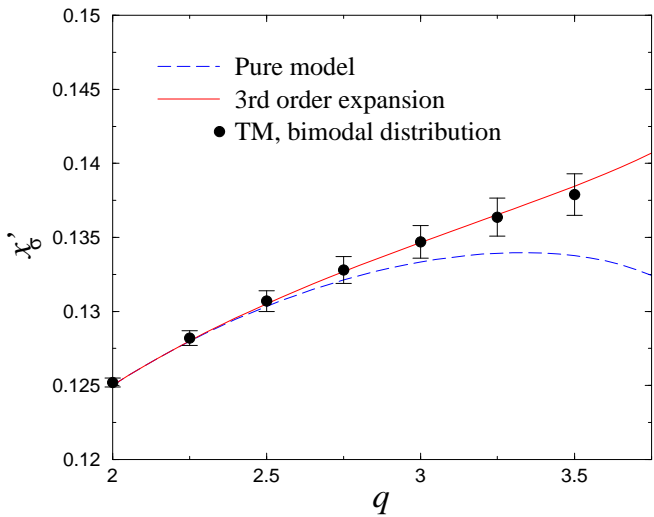,width=8.6cm}}
\vskip 0.2truecm
\caption{Scaling dimension of the order parameter (binary
disorder) compared to the third order
expansion of Dotsenko and 
co-workers~\protect\onlinecite{dotsenkopiccopujol95a}. The scaling dimension
corresponding to the pure model is shown for comparison.} 
\label{fig-xs(q)_p=1}
\end{figure}

From the exponential decay of the average correlation function, the exponent 
$x'_\sigma (q)$ is deduced and presented as a function of 
$q$ for the case of a binary disorder 
in Fig.~\ref{fig-xs(q)_p=1}. Although these results are not new, since the
same type of curve was reported by Cardy and Jacobsen 
in Ref.~\onlinecite{cardyjacobsen97},
the agreement with our results confirms the reliability of the averaging 
procedure.
The results are compared to the third order expansion
of Dotsenko {\it et al}~\cite{dotsenkopiccopujol95a} in 
Eq.~(\ref{eq:x_s-dotsenko}). The data themselves
are given in table~\ref{tab-x_ave-bin}.
The agreement is extremely good
especially in the region where the expansion is supposed to be valid when $q$
is not too far from the Ising model value $q=2$.

\vbox{
\begin{table}
\caption{Comparison of the numerical results for the magnetic scaling dimension
(bimodal probability distribution) $x'_\sigma(q)$ with the third order
expansion of Dotsenko and 
co-workers~\protect\onlinecite{dotsenkopiccopujol95a}. The 
error bars systematically contain the analytical value.
}
\vskip 0.2truecm
\begin{tabular}{lll}
$q$ & \multicolumn{2}{c}{$x'_\sigma$}\\
\cline{2-3}
& Expansion~(\protect\ref{eq:x_s-dotsenko}) & TM result \\ \hline
2 & 0.12500 & 0.1252(3)\\
2.25 & 0.12800 & 0.1282(5)\\
2.5 & 0.13051 & 0.1307(7)\\
2.75 & 0.13269 &  0.1328(9)\\
3. & 0.13465 &  0.1347(11)\\
3.25 & 0.13653 &  0.1364(13)\\
3.5 & 0.13845 &  0.1379(14)\\
\end{tabular}
\label{tab-x_ave-bin}\end{table}
}

Even close to the marginally irrelevant case of
the Ising model where logarithmic corrections are known to be present for
some quantities, we note that the numerical data are quite satisfactorily 
in agreement with
the perturbative results. This
is due to the absence of logarithmic corrections for the {\it average}
correlation function at $q=2$, and we will see
that this observation is no longer true in the following study of other 
moments.

\section{Tests of universality}
\label{sec:results}

\subsection{Universality of the average behaviour}
\label{subsec:Univ}

The question of universality in random systems is not yet solved, especially
when frustration occurs, like in random fields or spin 
glasses~\onlinecite{sourlas98}.
We address in this section the question of the influence of the
particular shape of the probability distribution of exchange couplings
on the universality class. There are still subsisting doubts concerning 
universality since
incompatible estimations (taking error bars into account) 
of the critical exponent $x'_\sigma$ were 
obtained with different probability 
distributions. For example at $q=8$, the choice of a bimodal probability 
distribution led to 0.151(4)~\onlinecite{picco98} (FSS)
or 0.1505(3)~\onlinecite{chatelainberche99} (conformal invariance),
 while a continuous
distribution gave 0.161(3)~\onlinecite{olsonyoung99} (FSS).

In this section, we show that the discrepancy is simply due to crossover effects
but does not imply absence of universality.
Following the methodology of the
previous sections, the disordered fixed point regime is located at the maximum
of the central charge and the scaling dimension $x'_{\sigma}$ of the average
spin-spin correlation functions is estimated. The results of 
Fig.~\ref{fig-xs(q)_p=1} are considered as a reference and the same 
method is applied to
ternary, quaternary and continuous distributions and is shown to lead to
identical critical exponents, within error bars, to those of a binary
distribution.

\subsubsection{Ternary distribution}
The ternary probability distribution is defined by
	\begin{equation}
	{\cal P}(K_{ij})={1\over 3}\left[
	\delta(K_{ij}-K_0)+\delta(K_{ij}-K)+\delta(K_{ij}-rK)\right]
	\label{eq7}\end{equation}
with the self-duality condition
	\begin{equation}
	[\exp (K_c(r))-1][\exp (rK_c(r))-1]=
	[\exp (K_0)-1]^2
	=q,
	\label{eq8}\end{equation}
where $K_0=\ln (1+\sqrt q)$ is the critical coupling of the pure system.
The homogeneous system corresponds to the value $r=1$.
We tried different kinds of interpolation procedures for the free energy, 
as in Eq.~(\ref{fit-c}), but always found a monotonic variation
 of the corresponding central charge
 with respect to the disorder amplitude $r$.
The absence of maximum might be the sign of the presence of strong
corrections, possibly due to the fact that one third of the exchange couplings
keeps their pure value $K_0$ even in the infinite-disorder limit.
Nevertheless, we present in the table~\ref{table4} the magnetic exponent as
extracted from the average spin-spin correlation functions for different values
of $r$.

\vbox{
\begin{table}
\caption{Scaling dimension $x'_\sigma$ of the average order parameter for the
$q=3$ Potts model with a ternary distribution compared to the results for a
binary distribution at the optimal disorder amplitude 
and for the pure model.}
\vskip 0.2truecm
\begin{tabular}{lll}
Distribution & disorder amplitude & $x'_\sigma(3)$\cr\hline
Pure   &  $r=1$		& $0.1333$      \cr\hline
Binary &  $\bf r^*\simeq 5.363$ & $\bf 0.1347(11)$  \cr \hline    
Ternary&  $r=2$ 	& $0.1339(3)$   \cr
       &  $r=4$         & $0.1343(6)$   \cr
       &  $\bf r=5.363$     & $\bf 0.1344(8)$   \cr
       &  $\bf r=8$         & $\bf 0.1344(10)$  \cr
       &  $r=12$        & $0.1343(13)$  \cr
       &  $r=20$        & $0.1341(15)$  \cr
\end{tabular}
\label{table4}\end{table}
}

For strong disorder, the scaling dimension $x'_\sigma$, as presented in the
table~\ref{table4}, shows a plateau with a value compatible within error bars
with that of the binary distribution, but the agreement is not yet conclusive,
since the effective central charge was not found to display a clear maximum.

\subsubsection{Quaternary distribution}
The quaternary probability distribution is defined by
	\begin{eqnarray}
	{\cal P}(K_{ij})&={1\over 4}\left[
	\delta(K_{ij}-K)+\delta(K_{ij}-rK)\right.      	 \nonumber\\
	&\left.+\delta(K_{ij}-K')+\delta(K_{ij}-r^2K')\right]
	\label{eq1}\end{eqnarray}
where the four equi-probable exchange couplings $K$, $rK$, $K'$
and $r^2K'$ are related by
	\begin{eqnarray}
	&&[\exp(K_c(r))-1][\exp(rK_c(r))-1]\nonumber\\
	&&=[\exp(K'_c(r))-1][\exp(r^2K'_c(r))-1]=q
	\label{eq2}\end{eqnarray}
at the self-dual point of the model. The value $r=1$ corresponds
to the pure system and the limit $r\rightarrow +\infty$ to a percolative
regime.

\begin{figure}[ht]
\centerline{\psfig{file=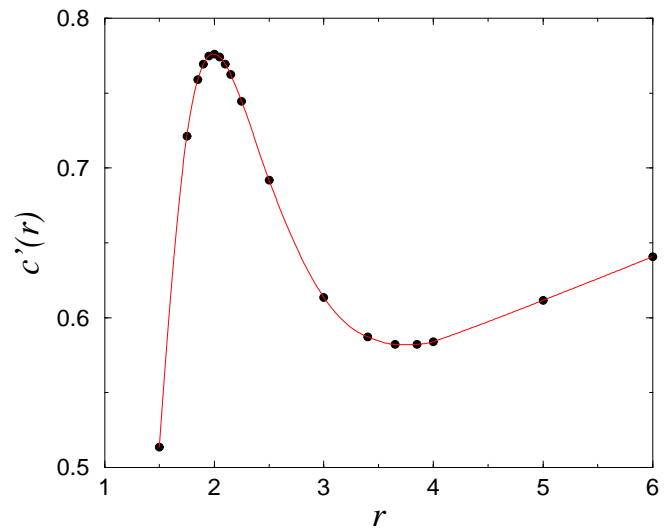,width=8.6cm}}
\vskip 0.2truecm
\caption{Behaviour of the central charge as a function of the disorder 
amplitude $r$ for the quaternary distribution~(\protect\ref{eq1}) at $q=3$.
The dotted curve is a guide for the eyes.} 
\label{fig1} \end{figure}

As seen on figure Fig.~\ref{fig1}, the central charge presents both
a maximum at $r^*\simeq 2.000$ and a ``minimum'' at $r_*\simeq 3.763$. According to
Zamolodchikov's $c$-theorem in the case of non-unitary theories, the latter
case should correspond to an instable fixed point while the former
situation is likely
to be the disordered
fixed point.

In table~\ref{table1}, we collect the scaling dimensions of the
average spin-spin correlation functions at these two fixed points and,
for comparison, those of the pure model and of the disordered system 
with a binary distribution.


\vbox{
\begin{table}
\caption{Scaling dimensions $x'_\sigma$ ($q=3$) of the average 
order parameter for the
two extrema ($r^*$ (written in bold face) and $r_*$) of the central 
charge with a quaternary
distribution compared to the results for the binary case and for
the pure model.}
\vskip 0.2truecm
\begin{tabular}{lll}
Distribution & disorder amplitude & $x'_\sigma(3)$\cr\hline
Pure   &  $r=1$        & $0.1333$      \cr\hline
Binary &  $\bf r^*\simeq 5.363$ & $\bf 0.1347(11)$  \cr\hline    
Quaternary   &  $\bf r^*\simeq 2.000$    & $\bf 0.1343(6)$   \cr      
    &  $r_*=3.763$    & $0.0097(12)$  \cr
\end{tabular}
\label{table1}\end{table}
}

Inspection of the results in table~\ref{table1}
reveals  that compatible exponents are obtained at the maximum 
of the central charge with
both probability distributions. These data also confirm once again that the
disorder amplitude $r$ plays an essential role~\onlinecite{picco98}, since
the value at $r_*$ is definitely excluded by the perturbative result. 

\subsubsection{Continuous distribution}
Recently, Olson and Young~\onlinecite{olsonyoung99} proposed  
slightly different numerical estimations
of the critical exponents of the average magnetization and of its first
moments, compared to other independent 
studies~\onlinecite{palagyichatelainbercheigloi99,chatelainberche99,lewis99}). 
They used an interesting 
continuous probability
distribution of exchange couplings that they claimed to be less sensible to
crossover effects. We show in the following that the randomness
amplitude chosen in the simulations of Olson and Young 
is not the optimal one, since it is
not strong enough to
reach the disordered fixed point. This observation may be at
the origin of  the slight 
discrepancy between the extrapolated values of the exponents.

\begin{figure}[ht]
\centerline{\psfig{file=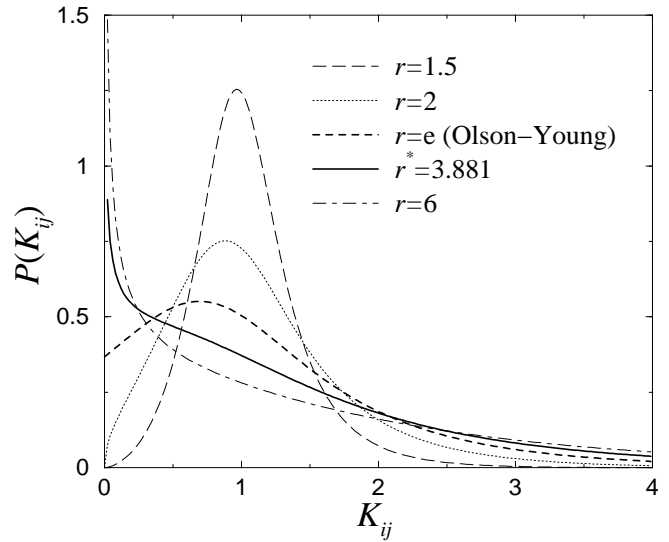,width=8.6cm}}
\vskip 0.2truecm
\caption{Probability distribution of exchange couplings ${\cal P}(K_{ij})$ for
disorder amplitudes $r$ in the range $[1.5;6]$.} 
\label{fig2} \end{figure}

The continuous probability distribution used by Olson and Young 
is generalized by introduction of
a parameter $r={\rm e}^\lambda$ which controls the strength of randomness. 
Following their notation,
we have chosen the distribution
	\begin{equation}
	{\cal P}(y_{ij})={1\over \cosh {y_{ij}\over \lambda}}
	\label{eq3}\end{equation}	
where
	\begin{equation}
	{\rm e}^{y_{ij}}={{\rm e}^{K_{ij}}-1\over\sqrt q}.
	\label{eq4}\end{equation}
Self-duality is ensured by the parity of the distribution ${\cal P}(y_{ij})$.
The definition of the disorder amplitude $r$ is such that
$r=1$ corresponds again to the pure system and $r={\rm e}$ to the Olson-Young distribution.
The probability distribution of exchange couplings $K_{ij}$ is given by
	\begin{equation}
	{\cal P}(K_{ij})=
	2{q^{1/2\lambda}\over \pi \lambda}{{\rm e}^{K_{ij}}({\rm e}^{K_{ij}}
	-1)^{1/\lambda-1}
	\over q^{1/\lambda}+({\rm e}^{K_{ij}}-1)^{2/\lambda}}
	\label{eq5}\end{equation}
and can be generated by the formula
	\begin{equation}
	K_{ij}=\ln\left(1+\sqrt q\tan^\lambda{\pi x\over 2}\right)
	\label{eq6}\end{equation}
if $x\in[0;1[$ is a uniformly distributed random variable.
Examples of probability distributions at different disorder amplitudes are
shown in Fig.~\ref{fig2}.


\begin{figure}[ht]
\centerline{\psfig{file=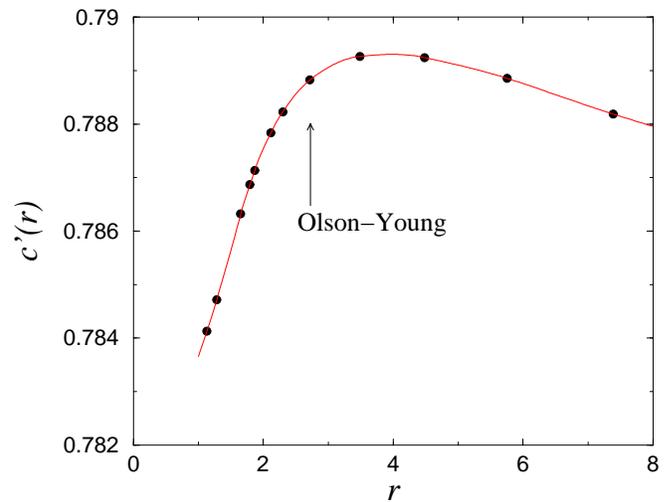,width=8.6cm}}
\vskip 0.2truecm
\caption{Central charge of the $q=3$ random bond Potts model for the continuous
distribution Eq.~(\ref{eq5}) with respect to the disorder amplitude $r$.} 
\label{fig3} \end{figure}


The maximum of the central charge is found in Fig.~\ref{fig3} at 
the amplitude of disorder $r^*\simeq
3.881$. 
The scaling dimension of the average order parameter, obtained at $r^*$,
is given in table~\ref{table3}  and compared to other disorder amplitudes.
Again, there is a convincing agreement with the results obtained with the
binary distribution.

With the continuous distribution at $r={\rm e}$, Olson and Young measured
slightly higher values for different values of $q$. We propose here a possible
explanation of the small discrepancy: 
It was shown that the larger the number of state $q$ of the Potts
model the stronger the disorder should be to reach the disordered fixed point
regime~\onlinecite{picco98,chatelainberche99,jacobsenpicco99}. Thus, for  $q\ge 3$
 random bond Potts models,
the {\it ideal} disorder amplitude should become larger and larger, far from
the value $r={\rm e}\simeq 2.718$ implicitly used by Olson and Young.
On the other hand, since in the weak disorder regime the average 
exponent is 
continuously growing with the 
disorder amplitude
$r$,
a too weak randomness cannot 
be the explanation
of the slightly too high values for the exponents obtained by Olson and Young.
A possible origin of the small disagreement can be found in the ensemble 
average: If the
number of disorder realizations is too small, the average
behaviour will give an exponent closer to the typical one, 
and thus too large.

\vbox{
\begin{table}
\caption{Scaling dimension $x'_\sigma$ of the average order parameter for the
$q=3$ random bond Potts model with a continuous distribution at the optimal
disorder amplitude (written in bold face) and other amplitudes as well,
compared to the results obtained in 
the binary case and for the pure model.}
\vskip 0.2truecm
\begin{tabular}{lll}
Distribution & disorder amplitude & $x'_\sigma(3)$\cr\hline
Pure   		&  $r=1$        & $0.1333$      \cr\hline
Binary 		&  $\bf r^*\simeq 5.363$ & $\bf 0.1347(11)$  \cr \hline    
Continuous      &  $r=1.133$    & $0.1338(1)$   \cr      
	        &  $r={\rm e}$  & $0.1340(9)$   \cr      
       		&  $\bf r^*\simeq 3.881$    & $\bf 0.1344(13)$  \cr      
       		&  $r=7.389$    & $0.1348(17)$   \cr      
\end{tabular}
\label{table3}\end{table}
}

\subsection{Replica Symmetry}
\label{subsec:RS}

The question of a possible breaking of replica symmetry~\onlinecite{mezard} in disordered
systems is very controversial and far from being settled, especially
in spin glasses (see e.g 
Refs.~\onlinecite{marinariparisiruizlorenzoritort96,marinarietal98a,bokilbraydrosselmoore98,marinarietal98b}).
In the context of disordered Potts ferromagnets, the question was first asked
by Dotsenko et al~\onlinecite{dotsenkodotsenkopicco97}. 
The Hamiltonian~(\ref{Ham}) is rewritten
\begin{eqnarray}
	-\beta{\cal H}[\sigma]&=&K_0\sum_{(i,j)}\delta_{\sigma_i,\sigma_j}+
	\sum_{(i,j)}(K_{ij}-K_0)\delta_{\sigma_i,\sigma_j}\nonumber\\
	&=&-\beta{\cal H}_0[\sigma]-\beta{\cal H}'[\sigma]
\label{eq-hamHH'}
\end{eqnarray}
where the deviation from the pure system can be written in the continuum limit
\begin{equation}
	-\beta{\cal H}'\sim\int\tau(x)\epsilon(x){\rm d}^2x
\label{eq-hamH'cont}
\end{equation}
with $\tau(x)=K(x)-K_c$. 
The average free energy $\overline{F}=-k_BT\overline{\ln Z}$ can be obtained
using the identity $\overline{\ln Z}=\lim_{n\to 0}\frac{\overline{Z^n}-1}{n}$
by introduction of $n$ identical copies (labelled by an index $a$) 
of the model, coupled by their energy densities. After integration over a
Gaussian probability distribution centered on the value $\tau_0$ and with
variance $\Delta$, one is led to:
\begin{eqnarray}
	\overline{Z^n}&=&{\rm Tr\ \!}\exp\left[
	-\beta\sum_a {\cal H}_0^{(a)}
	+\tau_0\int{\rm d}^2x\sum_a
	\epsilon_a(x)\right.\nonumber\\
	&+&\left. \Delta\int{\rm d}^2x\sum_{a\not=b}
	\epsilon_a(x)\epsilon_b(x)
	\right]
\label{replicas}
\end{eqnarray}
The first term governed by the average coupling $\tau_0$ produces a shift in the
critical temperature, while the second term couples the replicas with each other. 
In a replica symmetric scenario, the couplings $\Delta$ between the different copies
of the model are identical, while in a Replica Symmetry Breaking scheme, these
couplings are Parisi matrices and can take different values $\Delta_{ab}$.
Treated as a
perturbation this coupling term leads to different fixed point structures and
finally to different scaling dimensions for the moments of the correlation 
functions.
In order to test between 
Replica Symmetry  and Replica Symmetry Breaking
 schemes, Dotsenko et al performed a second order expansion of the exponent of the 
second moment of the spin-spin correlation function decay in both 
cases (Equations~(\ref{eq:RS}) and (\ref{eq:RSB})). 
Previous MC simulations have been performed at $q=3$ but were not completely
conclusive, although in favour of Replica Symmetry: The perturbation expansion
leads to $x'_{\sigma^2}(3)=0.1176$ and $x''_{\sigma^2}(3)=0.1201$ according
to Eq.~(\ref{eq:RS}) and (\ref{eq:RSB}), while previous numerical results lead to
$0.113(1)$~\onlinecite{dotsenkodotsenkopicco97},
$0.1140(5)$~\onlinecite{lewis99}, 
$0.116(1)$~\onlinecite{palagyichatelainbercheigloi99} and
$0.119(2)$~\onlinecite{olsonyoung99}.

In this section, we report new conclusive results for different values of $q$.
Close to $q=2$, the proximity of the marginally irrelevant Ising FP will
surely alter the data, as a reminiscent effect of the logarithmic 
corrections present exactly at $q=2$ for the second moment. Too large values of
$q$ on the other hand are not very helpful in order to check perturbation
expansions which break down when one explores higher values of the expansion 
parameter (as given for example in table~\ref{table-max-c}). One thus has to
balance these two extreme situations and the comparison between numerical
data and perturbation results should be conclusive around $q=3$. 
The TM technique thus appears to be well adapted, since it is capable to deal
with non integer values of $q$.

\begin{figure}[ht]
\centerline{\psfig{file=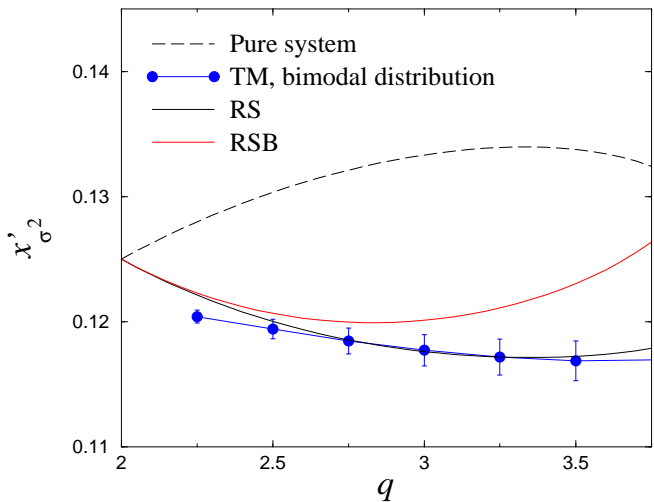,width=8.6cm}}
\vskip 0.2truecm
\caption{Exponent of the second moment of the spin-spin correlation function
as a function of the number of states of the disordered Potts model 
(binary disorder). The
comparison is done with Replica Symmetry  and Replica Symmetry Breaking
 scenarios~\protect\onlinecite{dotsenkodotsenkopicco97}. The agreement
 with the RS result is quite good around $q=3$. When $q$ is close to 2, the discrepancy can be attributed
to the weak relevance of disorder. We indeed used a simple exponential fit as can
be expected at a stable disordered FP, but at $q=2$, one knows from Ludwig's
results that logarithmic corrections must be added. These corrections can also
influence the vicinity of $q=2$ in a numerical approach.} \label{fig-xs2-vs-q}
\end{figure}

The comparison is shown in Fig.~\ref{fig-xs2-vs-q} for the bimodal 
probability distribution and the results are also
given in table~\ref{table-xs2-vs-q}. In the convenient domain for the test, 
around $q=3$, results are written in bold face.
 The agreement with Replica Symmetry
is quite convincing for the bimodal 
probability distribution. We note that with the 
continuous distribution of Eq.(\ref{eq5}),
we obtain also a very good value at $q=3$: $0.1173(14)$ and with the ternary distribution at $r=8$
we get $0.1182(12)$.

\vbox{
\begin{table}
\caption{Decay exponent of the second moment of the spin-spin correlation 
function compared to Replica Symmetry and Replica Symmetry Breaking
 expressions of Eqs.~(\ref{eq:RS}) and (\ref{eq:RSB}). The results 
 written in bold face correspond to the range of values of $q$ where the agreement
 is particularly satisfactory.\label{table-xs2-vs-q}}
\begin{tabular}{llll}
&\multicolumn{2}{c}{Perturbative results}&TM result\\
\cline{2-3}
	$q$ & $x'_{\sigma^2}$ & $x''_{\sigma^2}$ & Binary Disorder\\
	\hline
2.25	&0.12213	&0.12229	&0.1204(5)	\\	
2.5	&0.12002	&0.12067	&0.1194(8)	\\	
\bf 2.75	&\bf 0.11854	&0.11997	&\bf 0.1185(10)	\\	
\bf 3.	&\bf 0.11761	&0.12011	&\bf 0.1177(12)	\\
\bf 3.25 &\bf 0.11718	&0.12110	&\bf 0.1172(14)\\	
\bf 3.5	&\bf 0.11723	&0.12304	&\bf 0.1169(16)	\\
\hline
&&&Ternary Disorder	\\
\bf 3.&\bf 0.11761&0.12011& \bf 0.1182(12)\\
\hline
&&&Continuous Disorder	\\
\bf 3.&\bf 0.11761&0.12011& \bf 0.1173(14)\\
\end{tabular}
\end{table}
}

\subsection{Multifractality}
\label{subsec:Multi}

\begin{figure}[ht]
\centerline{\psfig{file=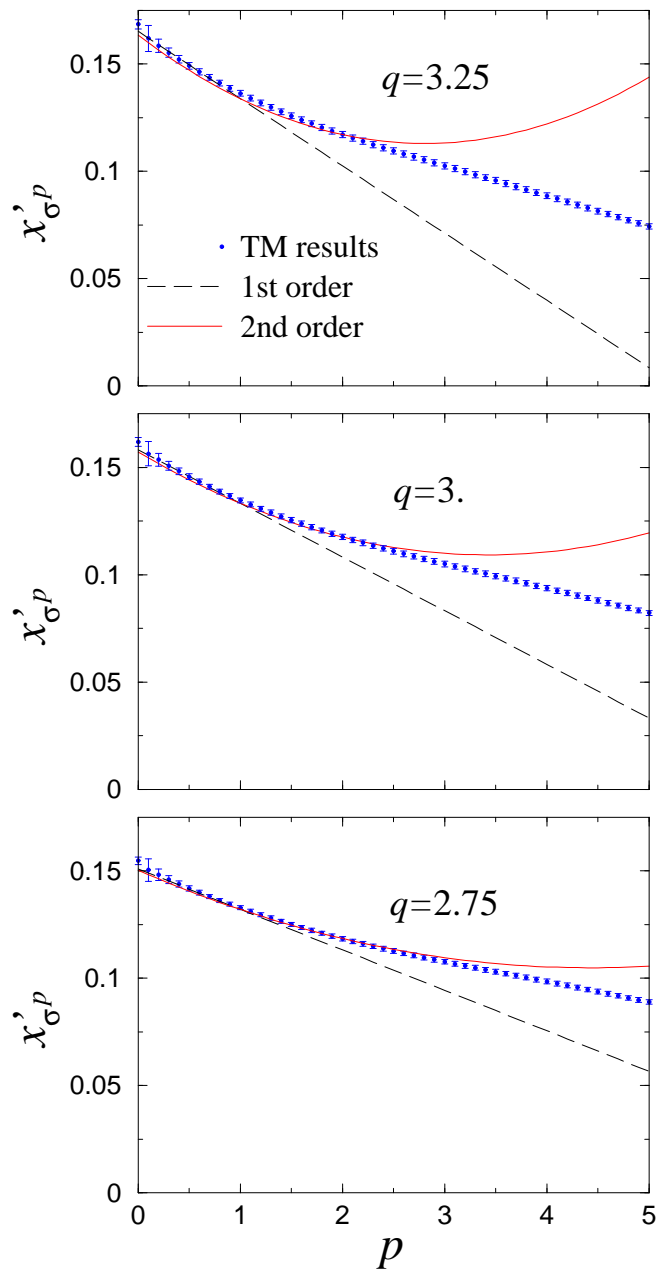,width=8.6cm}}
\vskip 0.2truecm
\caption{Comparison of the multifractal exponents (reduced moment of the
correlation function $\overline{\langle\sigma(0)\sigma(u)\rangle^p
}^{1/p}$) with the second order
expansion of Lewis in the RS scheme~\protect\onlinecite{lewis98} for different
values of $q$ indicated in the figure (bimodal probability distribution).} \label{fig-q=2.75_moments}
\end{figure}

The multiscaling behaviour of the spin-spin correlation functions is
noticeable in the $p-$dependent set of exponents of the reduced moments 
$\overline{\langle\sigma(0)\sigma(\rho)\rangle^p}^{1/p}$. The second moment
has already been computed when looking for Replica Symmetry, and it can
be generalized in the strip geometry using Eq.~(\ref{eq:G-ruban}).
We performed an exhaustive computation of 50 different moments in the range
$0\leq p\leq 5$ in the strip geometry, and the associated scaling dimensions followed from a
semi-log fit $\ln \overline{\langle\sigma(0)\sigma(u)\rangle^p}
$ vs $\ln u$, according to Eq.~(\ref{eq:G-ruban}), followed by an extrapolation
to $L\to\infty$. Examples for $q=2.75$, 3 and 3.25 are shown
in Fig.~\ref{fig-q=2.75_moments} (bimodal probability distribution) 
where the numerical results are also compared
to the first order expansion of Ludwig and to 
the second order
expansion in the RS scheme in Eq.~(\ref{eq:x_lewis}). The second order
result is clearly very good up to values of $p$ close to 3 and then breaks down
as already noticed by Lewis~\onlinecite{lewis99}. 

\begin{figure}[ht]
\centerline{\psfig{file=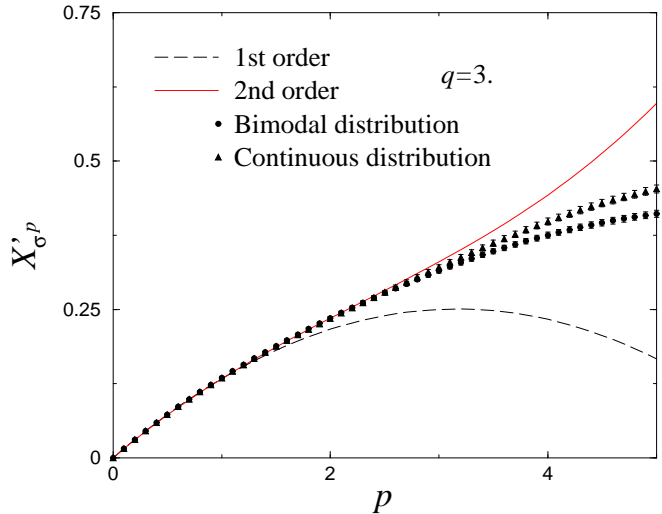,width=8.6cm}}
\vskip 0.2truecm
\caption{Comparison of the multifractal exponents (moment of
the correlation function $\overline{\langle\sigma(0)\sigma(u)\rangle^p}$) with the second order
expansion of Lewis in the RS scheme~\protect\onlinecite{lewis98} for
both the bimodal and the continuous probability distributions.} 
\label{fig-q=3_moments_px}
\end{figure}

An alternate presentation of the results (used e.g. by Ludwig~\onlinecite{ludwig90})
is given by the scaling dimension of the moment of the correlation function
itself, $\overline{\langle\sigma(0)\sigma(\rho)\rangle^p}$
(not the reduced function $\overline{\langle\sigma(0)\sigma(\rho)\rangle^p
}^{1/p}$). The scaling dimension is thus simply $px'_{\sigma^p}(q)$, hereafter
denoted by $X'_{\sigma^p}(q)$ by a simple extension of Ludwig's notation.
An example, with $q=3.$, is shown in Fig.~\ref{fig-q=3_moments_px} where
we have also shown the results obtained with the continuous probability distribution
at the optimal disorder amplitude. Once again, we find a promising agreement
between the numerical data and the perturbative result which confirms 
universality.

\begin{figure}[ht]
\centerline{\psfig{file=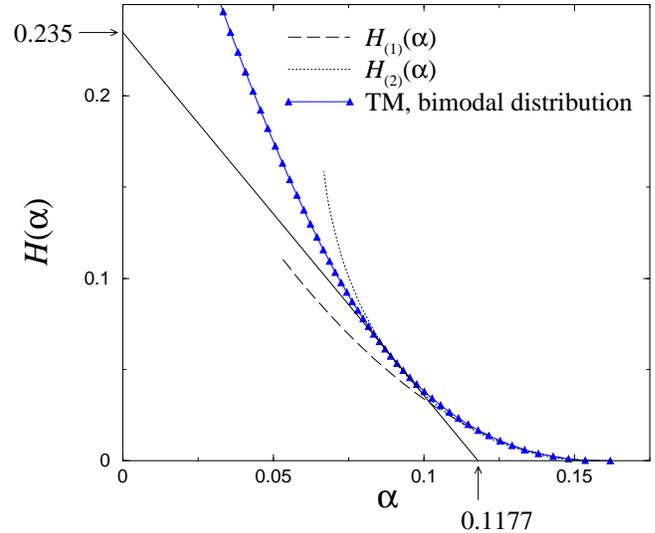,width=8.6cm}}
\vskip 0.2truecm
\caption{Universal function $H(\alpha)$ for different $q=3$. The functions
$H_{(1)}(\alpha)$ and $H_{(2)}(\alpha)$ 
deduced from Eq.~(\ref{eq:x_lewis}) at first or second order, respectively, 
are shown for 
comparison.} 
\label{fig-H-de-alpha-q=3}
\end{figure}

\begin{figure}[ht]
\centerline{\psfig{file=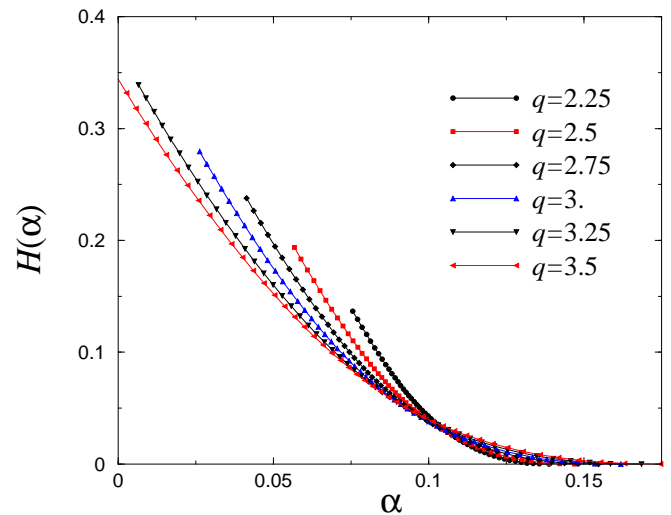,width=8.6cm}}
\vskip 0.2truecm
\caption{Universal function $H(\alpha)$ for different values of $q$
(bimodal probability distribution).} 
\label{fig-H-de-alpha-tsq}
\end{figure}

What is interesting in this latter presentation is the link with other
fields where multifractality is observed. One then usually introduces
a universal function, the multiscaling function $H(\alpha)$, which is simply
the Legendre transform of the set of independent scaling indexes 
$X'_{\sigma^p}(q)$. Setting ${\rm d}X'_{\sigma^p}(q)=\alpha{\rm d}p$, this
function is simply obtained by $H(\alpha)=X'_{\sigma^p}(q)-\alpha p$. The
geometrical
interpretation of this Legendre transform follows from the relation
$\frac{\partial H}{\partial \alpha}=-p$ where $\alpha$ is defined by
$\frac{\partial X'_{\sigma^p}(q)}{\partial p}=\alpha$. The scaling
dimension $x'_{\sigma^p}(q)$ is obtained on the plot of $H(\alpha)$ 
by the intercept of the tangent of slope
$-p$ with the abscissa axis. An example of multiscaling function $H(\alpha)$
deduced from the numerical data with the bimodal probability distribution 
is shown in Fig.~\ref{fig-H-de-alpha-q=3}
for $q=3$ and the line of slope $-2$, leading to the exponent 
of the second moment (See table~\ref{table-xs2-vs-q})
is also shown. 
For comparison, the function $H_{(k)}(\alpha)$ deduced from the $k-$th order 
(in $y_H$) perturbative
results of Ludwig ($k=1$)  and 
Lewis ($k=2$) (\ref{eq:x_lewis}) are also plotted.
Other values of $q$ are shown in Fig.~\ref{fig-H-de-alpha-tsq}.

\subsection{Correlation function probability distribution}
\label{subsec:prodist}

In Ref.~\onlinecite{ludwig90}, Ludwig presented a remarquable discussion of
the spin-spin correlation function probability distribution. He showed how all the relevant
information on the large distance behaviour is encoded in the multifractal function
$H(\alpha)$. In this section, we follow Ludwig's arguments and report a
numerical study of the correlation function probability distribution in the 
cylinder geometry.

\begin{figure}[ht]
\centerline{\psfig{file=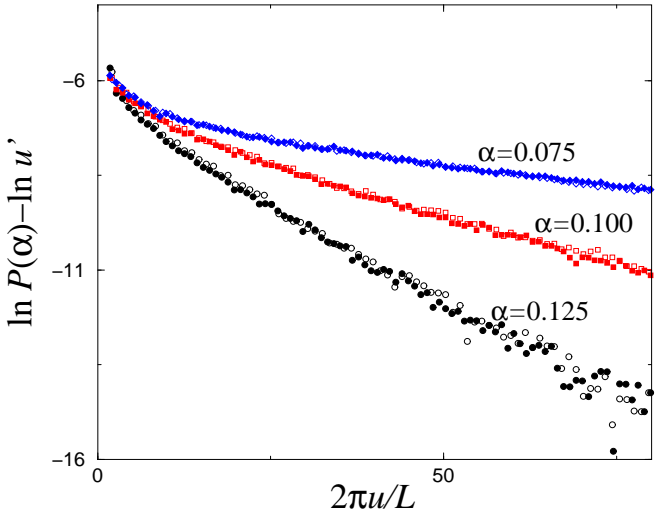,width=8.6cm}}
\vskip 0.2truecm
\caption{Behaviour of the probability distribution ${\cal P}(\alpha)$
as a function of the distance along the strip $2\pi u/L$.
The rescaled position along the strip is written $u'=2\pi u/L$.
Three fixed values of $\alpha$ are shown, the opened and filled symbols
respectively correspond to the 
strip widths $L=6$ and 7. A good collapse of the data at both sizes is observed,
but the behaviour displays a deviation from linearity at small
distances
($q=3$, binary distribution).} 
\label{fig-DistPr-alpha-1}
\end{figure}

According to the results of the previous section, the moments of the
spin-spin correlation function along the strip asymptotically behaves as follows:
\begin{equation}
	\overline{G^p(u)}\equiv\overline{\langle\sigma(0)\sigma(u)\rangle^p}
	\sim B_p{\rm e}^{-\frac{2\pi u}{L}X'_{\sigma^p}}
\label{eqProb1}
\end{equation}
and are defined in terms of the probability distribution ${\cal P}[G(u)]$:
\begin{equation}
	\overline{G^p(u)}=\int_0^1{\rm d} G(u){\cal P}[G(u)]G^p(u).
\label{eqProb2}
\end{equation}
Following Ludwig, we introduce the variable $Y(u)=-\ln G(u)$
and write $G^p(u)={\rm e}^{-pY(u)}$. Using the identity
${\cal P}[G(u)]{\rm d}G={\cal P}[Y(u)]{\rm d}Y$ and 
equations~(\ref{eqProb1}) and (\ref{eqProb2}), one obtains
\begin{equation}
	\int_0^\infty{\rm d} Y(u){\cal P}[Y(u)]{\rm e}^{-pY(u)}
	\sim B_p{\rm e}^{-\frac{2\pi u}{L}X'_{\sigma^p}}
\label{eqProb3}
\end{equation}
which leads to
the expression of the probability distribution by inverting the 
Laplace transform ($\delta>0$):
\begin{equation}
	{\cal P}[Y(u)]=\frac{1}{2{\rm i}\pi}
	\int_{\delta-{\rm i}\infty}^{\delta+{\rm i}\infty} 
	{\rm d} pB_p{\rm e}^{-\frac{2\pi u}{L}\left[X'_{\sigma^p}
	-\frac{Y(u)}{2\pi u/L}p\right]}.
\label{eqProb4}
\end{equation}
The amplitude $B_p$ is weakly depending on $p$.  Following Ludwig, it can be
rewritten as $B_p=\exp \left[{-\frac{2\pi u}{L}
\left(-\frac{\ln B_p}{2\pi u/L}\right)}\right]$,  but 
 can be forgetten, since it only introduces a small correction 
 when $2\pi u/L\to\infty$.
Let us define the function $h(p)=X'_{\sigma^p}
	-\frac{Y(u)}{2\pi u/L}p$.
In the large distance limit $2\pi u/L\to\infty$, the integral can be 
evaluated by the saddle-point approximation at the minimum $p_0$ of $h(p)$:
\begin{equation}
	\left(\frac{\partial}{\partial p}X'_{\sigma^p}\right)_{p_0}=
	\frac{Y(u)}{2\pi u/L}
\label{eqProb5}
\end{equation}
Instead of $Y(u)$, we define the scaled variable 
$\alpha=\frac{Y(u)}{2\pi u/L}$, and the saddle point value at $p_0$ only depends
on this variable $h(p_0)=H(\alpha)$, where $H(\alpha)$ is nothing but the 
multifractal function
defined in the previous section. We thus obtain the probability distribution
\begin{equation}
	{\cal P}[Y(u)]\sim\exp\left[{-\frac{2\pi u}{L}H\left(\frac{Y(u)}
	{2\pi u/L}\right)}\right],
\label{eqProb6}
\end{equation}
or, using ${\cal P}[Y(u)]{\rm d}Y={\cal P}(\alpha){\rm d}\alpha$, 
\begin{equation}
	{\cal P}(\alpha)\sim\frac{2\pi u}{L}\exp\left[{-\frac{2\pi u}
	{L}H(\alpha)}\right].
\label{eqProb7}
\end{equation}
The multifractal function contains the essential information on the probability
distribution. In order to check this expression,  the value of 
$H(\alpha)$ at fixed $\alpha$ is extracted by fitting the probability 
distribution to
the expression
\begin{equation}
	\ln{\cal P}(\alpha)={\rm const}+\ln\frac{2\pi u}{L}-\frac{2\pi u}
	{L}H(\alpha).
\label{eqProb8}
\end{equation}
It is shown in Fig.~\ref{fig-DistPr-alpha-1} where the probability
distribution of the spin-spin correlation function was obtained
after collecting the results over $96\ 000$ disorder realizations
in 50 classes. The values of $H(\alpha)$
are slightly too large, compared to the results presented in the previous 
section.
We can indeed observe in Fig.~\ref{fig-DistPr-alpha-1} a deviation from
the linear behaviour which would be expected with these variables, and the
shorter the distance $u$, the larger the deviation.

\begin{figure}[ht]
\centerline{\psfig{file=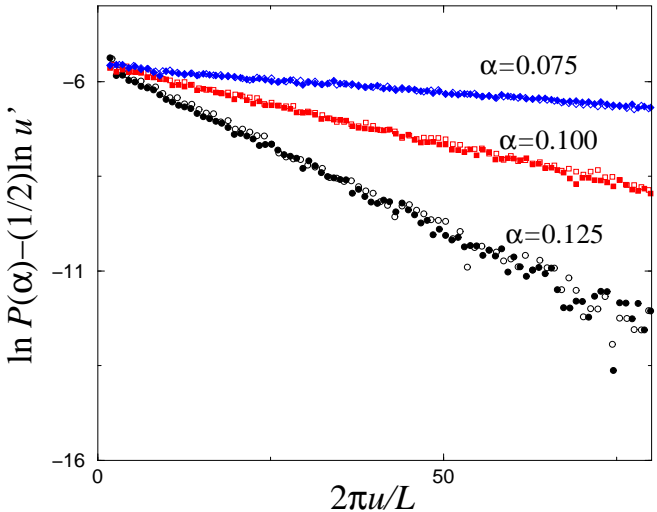,width=8.6cm}}
\vskip 0.2truecm
\caption{Same as Fig.~\protect\ref{fig-DistPr-alpha-1} accounting for the 
correction in Eq.~(\protect\ref{eqProb10}) close to the 
saddle-point approximation.} 
\label{fig-DistPr-alpha-2}
\end{figure}

\begin{figure}[ht]
\centerline{\psfig{file=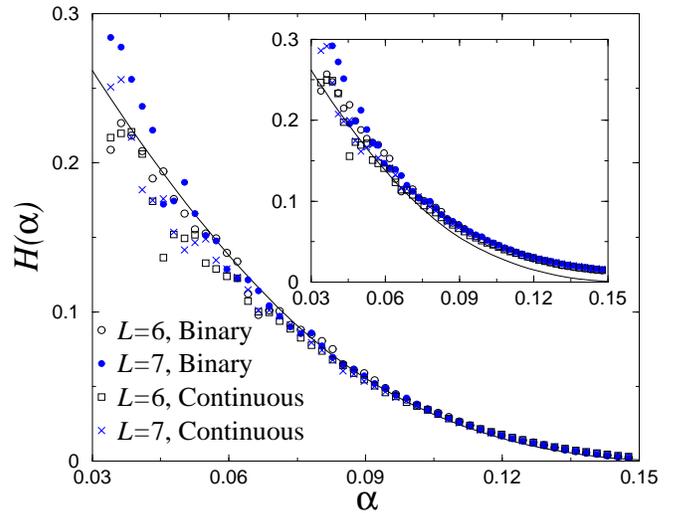,width=8.6cm}}
\vskip 0.2truecm
\caption{Multifractal function $H(\alpha)$ as it is deduced from the fit of the
probability distribution ${\cal P}(\alpha)$ to Eq.~(\protect\ref{eqProb10})
accounting for the 
correction near the saddle-point approximation.
It is compared to the results of Fig.~\protect\ref{fig-H-de-alpha-q=3} in
solid line. The insert shows slightly too large values of $H(\alpha)$
when deduced from
Eq.~(\protect\ref{eqProb8}).} 
\label{fig-DistPr-alpha-3}
\end{figure}

This could be due to
a correction to the leading behaviour given by the saddle-point 
approximation~\footnote{We mention here that
a possible correction to the saddle-point approximation has been suggested by Olson and Young in their study of
higher values of $q$.}.
If we expand the function $h(p)$ close to $p_0$, $h(p)\simeq H(\alpha)+\frac{1}
{2}h''(p_0)(p-p_0)^2$, with $h''(p_0)>0$ we obtain, instead of Eq.~(\ref{eqProb6}),
the following result for 
the probability 
distribution ${\cal P}[Y(u)]$~\onlinecite{debruijn}:
\begin{equation}
	{\cal P}[Y(u)]\sim\left(\frac{2\pi u}{L}\right)^{-1/2}
	\exp\left[{-\frac{2\pi u}
	{L}H
	\left(\frac{Y(u)}
	{2\pi u/L}\right)}\right]
\label{eqProb9}
\end{equation}
and a correction appears in ${\cal P}(\alpha)$:
\begin{equation}
	\ln{\cal P}(\alpha)-\frac{1}{2}\ln\frac{2\pi u}{L}={\rm const}-\frac{2\pi u}
	{L}H
	(\alpha).
\label{eqProb10}
\end{equation}
This is shown in Fig.~\ref{fig-DistPr-alpha-2} where a linear behaviour is
now obtained in the whole range of values of $u/L$. 
A linear fit in the
coordinates of Fig.~\ref{fig-DistPr-alpha-2} gives the value of the 
multifractal function $H(\alpha)$ which
can be compared to the results of Fig.~\ref{fig-H-de-alpha-q=3} 
obtained in section~\ref{subsec:Multi}. This fit is performed for all values
of $0.034< \alpha<0.15$ for $q=3$ for the cases of the bimodal probability distribution
and of the continuous distribution. The results are shown in 
Fig.~\ref{fig-DistPr-alpha-3}.

This latter figure shows that the correlation function probability distribution
is entirely determined by the universal multifractal function $H(\alpha)$.

\section{Conclusions}
\label{sec:conclusion}

In this paper, we have investigated the critical behaviour of the moments of
the spin-spin correlation functions of two-dimensional random bond Potts
ferromagnets using Transfer Matrix techniques and conformal methods. New features
of our present work are the following. 
\begin{description}
\item[i)]  As far as we know, universality of the critical behaviour of the
moments of the correlation function
was checked for the first time in such systems\footnote{We mention here that
a MC study in the first-order regime of the pure model ($q=5$) was
recently reported wgere random bond disorder and dilution were found
to belong to the same universality 
class~\protect\onlinecite{paredesvalbuena99}.}. This statement follows from
the numerical evidence that the scaling dimension of the average spin-spin
correlation function, as well as those of higher moments or typical behaviour
do not depend on the details of the probability distribution, provided that
the computations are performed at the disordered fixed point given by the
maximum condition of the central charge. The exponent of the average correlation
function
is furthermore in very good agreement with theoretical results using
perturbative conformal field theory available in 
the literature.
\item[ii)] The question of a possible breaking of Replica Symmetry is also
considered and the numerical data strongly support the absence of Replica Symmetry
Breaking. The problem is investigated through the comparison of the scaling dimension
of the second moment of the correlation function, which is compared to 
perturbative results obtained within both schemes.
Although previous numerical results  (using Monte Carlo
simulations) which led to similar conclusions were already reported at $q=3$, 
we believe that our study is conclusive, since it extends the work to 
other non integer values
of $q$.
\item[iii)] The multiscaling behaviour of the spin-spin correlation function
is investigated. The exponential decay of the moments of the correlation 
functions
along very long strips is used to deduce numerically the corresponding critical
exponents. These dimensions continuously depend on the moment order, as a
consequence of multifractality. They are furthermore very weakly dependent
of the probability distribution. At low moment order, the numerical 
results are furthermore in
agreement with a perturbative result obtained within the Replica Symmetry 
scheme. When the moment order increases, a discrepancy is observed, resulting 
from 
the lack of validity of the perturbation expansion, and possibly of a numerical
determination which becomes less precise. The multifractal function, given
by the Legendre transform of the set of independent scaling dimensions is also
computed for different values of $q$. It is shown that this universal
function completely determines the shape of the correlation function probability
distribution.
\end{description}
The main result of this paper is probably to show that universality 
in random systems has to be understood in the sense of a critical behaviour
which does not depend on the choice of the probability distribution (this is only
true up to some extent, since special distributions which do not obey the 
central limit theorem, like Levy flights, would certainly lead to different
results). By critical behaviour, here we mean the behaviour of all the moments
of a physical quantity, entirely contained in the multifractal function or
the correlation function probability
distribution
but we want to stress that lack of self-averaging does not imply absence of
universality.

We also note that logarithmic corrections were recently reported in the disconnected
energy-energy correlations function~\onlinecite{cardy99} and that disorder was 
shown to induce non-vanishing cross-correlations between spin-spin and energy-energy
moments~\onlinecite{daviscardy99}. The multiscaling of energy correlations 
has been studied
very recently by J.L. Jacobsen~\onlinecite{jacobsen99}.


\acknowledgments
We would like to thank M.A. Lewis and I. Campbell for stimulating discussions. 
The computations  were performed on the 
{\it SP2} at the CNUSC in Montpellier under project No. C990011, 
and the {\it Power Challenge Array} at the CCH in Nancy. 


\newcommand{\Name}[1]{\rm  #1,}
\newcommand{\And}{\ and\ }
\newcommand{\Review}[1]{\it  #1\rm}
\newcommand{\Vol}[1]{\bf  #1\rm,}
\newcommand{\Year}[1]{\rm  (#1)}
\newcommand{\Page}[1]{\rm  #1}
\newcommand{\Book}[1]{\it  #1\rm}

 \def\JPC{J. Phys. C: Solid State Phys.}
 \def\PRB{\it Phys. Rev. B}
 \def\PRE{\it Phys. Rev. E}
 \def\PRL{\it Phys. Rev. Lett.}
 \def\JSP{\it J. Stat. Phys.}
 \def\JPF{\it J. Phys.  I France}
 \def\ZPB{\it Z. Phys. B}

\def\paper#1#2#3#4#5{#1, #3 {\bf #4}, \rm #5 (#2).}

\end{multicols}
\end{document}